\documentclass[preprint,3p,times,twocolumn]{elsarticle}

\usepackage{amsmath}
\usepackage{amssymb}
\usepackage{amsthm}
\usepackage{amscd}
\usepackage{bbm}
\usepackage{mathrsfs}
\usepackage[mathscr]{eucal}
\usepackage{tikz-cd}

\newtheorem{thm}{Theorem}[section]

\theoremstyle{remark}

\newtheorem{example}[thm]{Example}

\usepackage{amssymb}
\journal{Encyclopedia of Mathematical Physics}

\begin{document}

\begin{frontmatter}

%% Title, authors and addresses

%% use the tnoteref command within \title for footnotes;
%% use the tnotetext command for theassociated footnote;
%% use the fnref command within \author or \address for footnotes;
%% use the fntext command for theassociated footnote;
%% use the corref command within \author for corresponding author footnotes;
%% use the cortext command for theassociated footnote;
%% use the ead command for the email address,
%% and the form \ead[url] for the home page:
%% \title{Title\tnoteref{label1}}
%% \tnotetext[label1]{}
%% \author{Name\corref{cor1}\fnref{label2}}
%% \ead{email address}
%% \ead[url]{home page}
%% \fntext[label2]{}
%% \cortext[cor1]{}
%% \affiliation{organization={},
%%             addressline={},
%%             city={},
%%             postcode={},
%%             state={},
%%             country={}}
%% \fntext[label3]{}

\title{Topological Semimetals}

%% use optional labels to link authors explicitly to addresses:
%% \author[label1,label2]{}
%% \affiliation[label1]{organization={},
%%             addressline={},
%%             city={},
%%             postcode={},
%%             state={},
%%             country={}}
%%
%% \affiliation[label2]{organization={},
%%             addressline={},
%%             city={},
%%             postcode={},
%%             state={},
%%             country={}}

\author{Guo Chuan Thiang}
\affiliation{organization={Beijing International Center for Mathematical Research, Peking University},
%%             addressline={},
             city={Beijing},
%%             postcode={},
%%             state={},
            country={China}}

\begin{abstract}
%% Text of abstract
We review the differential topology underlying the topological protection of energy band crossings in Weyl semimetals, and how they lead to the experimental signature of surface Fermi arcs.
\end{abstract}

%%Graphical abstract
%\begin{graphicalabstract}
%\includegraphics{grabs}
%\end{graphicalabstract}

%%Research highlights
%\begin{highlights}
%\item Research highlight 1
%\item Research highlight 2
%\end{highlights}

\begin{keyword}
%% keywords here, in the form: keyword \sep keyword
Weyl semimetals\sep Differential topology\sep Fermi arcs\sep Topological states\sep Monopoles
\sep Spectral flow \sep Dirac strings \sep Chern class
%% PACS codes here, in the form: \PACS code \sep code

%% MSC codes here, in the form: \MSC code \sep code
%% or \MSC[2008] code \sep code (2000 is the default)

\end{keyword}

\end{frontmatter}

\newcommand{\CC}{\mathbb{C}}
\newcommand{\NN}{\mathbb{N}}
\newcommand{\RR}{\mathbb{R}}
\newcommand{\PP}{\mathbb{P}}
\newcommand{\TT}{\mathbb{T}}
\newcommand{\ZZ}{\mathbb{Z}}

%% \linenumbers

%% main text

\section{Introduction}
Weyl semimetals are crystalline materials, usually in 3D, whose energy band structures have crossings that enjoy a certain topological protection. They are of great theoretical interest because of their analogy with the elusive Weyl fermion and the chiral anomaly. In the condensed matter physics context, these protected crossings are argued to occur in cancelling pairs. More significantly, they are predicted to have unusual \emph{Fermi arc} states localized near the material boundary. Remarkably, Fermi arcs were discovered in 2015 \cite{Lv1, Xu1, Xu2}, and many other experiments, see \cite{Lv} for an overview. There already exist several reviews on this subject from the physics perspective \cite{AMV,Lv,Gao}. This article provides a mathematical perspective, with emphasis on the global differential topology aspects of general band crossings, which are essential to the Weyl semimetal phenomenon.

\section{Fourier transform and topology}\label{sec:Fourier}
Let $V$ be a $d$-dimensional real vector space, to be thought of as the additive group of translational symmetries of the underlying affine ``position space''. We refer to the dual vector space $V^*={\rm Hom}(V,\RR)$ as \emph{momentum space}. 

\subsection{Brillouin torus}
Let $\Gamma\subset V$ be a lattice of rank $d$, i.e., $\Gamma$ is a discrete subgroup generated by $d$ linearly independent vectors in $V$.
Each $p\in V^*$ determines a character (one-dimensional unitary representation) of the additive group $V$,
\[
\chi_p:V\to{\rm U}(1),\qquad x\mapsto e^{ip(x)}.
\]
By restriction, $\chi_p$ can also be regarded as a character of the lattice subgroup $\Gamma\subset V$, but now there is redundancy in the label $p$. Namely, there is a discrete subgroup $\Gamma^\perp\subset V^*$ which annihilates $\Gamma$,
\[
\Gamma^\perp:=\{p\in V^*\,:\,\chi_p(x)\equiv e^{ip(x)}=1,\;\;\forall x\in\Gamma\},
\]
called the \emph{reciprocal lattice}. The quotient group
\[
\widehat{\Gamma}=V^*/\Gamma^\perp={\rm Hom}(\Gamma,{\rm U}(1))
\]
is called the \emph{Brillouin torus} of \emph{quasimomenta}. Note that $V$ and $V^*$ are naturally manifolds (they are Lie groups), so $\widehat{\Gamma}$ is actually a smooth manifold. 

A choice of lattice basis gives an identification $\Gamma\cong\ZZ^d$, as well as $V\cong\RR^d$. The dual basis identifies $V^*\cong\RR^d$, and $\Gamma^\perp\cong(2\pi \ZZ)^d$. Then $\widehat{\Gamma}$ is identified as $\RR^d/(2\pi\ZZ)^d\cong\TT^d$, and may be labelled by $d$-tuples $(e^{i\theta_1},\ldots, e^{i\theta_d})$ of unit complex numbers, with composition in $\widehat{\Gamma}$ corresponding to addition of the $\theta_j$ modulo $2\pi$. One usually takes the phases $\theta_j$ to lie in $[-\pi,\pi]_{-\pi\sim\pi}$.  Although $(\theta_1,\ldots, \theta_d)$ looks like a $d$-component ``vector'', it should really be viewed as \emph{local coordinates} for the manifold $\widehat{\Gamma}$, adapted to a choice of lattice basis.

\subsection{Fourier transform}
The Fourier transform is
\begin{align*}
\mathcal{F}:\ell^2(\Gamma)&\to L^2(\widehat{\Gamma})\\
\psi &\mapsto \left(\widehat{\psi}:\chi\mapsto\sum_{x\in\Gamma}\psi(x)\chi(x)\right),
\end{align*}
and is unitary when the Brillouin torus $\widehat{\Gamma}$ is given the translation-invariant Haar measure, normalized to $(2\pi)^{-d}$ times the Lebesgue measure in terms of the local coordinates $\vartheta=(\theta_1,\ldots,\theta_d)$.

The Hilbert space $\ell^2(\Gamma)$ admits the obvious representation of $\Gamma$ by translation. A lattice basis gives a generating set of translation operators $S_j, j=1,\ldots,d$ acting on $\ell^2(\Gamma)\cong\ell^2(\ZZ^d)$. Under conjugation by $\mathcal{F}$, $S_j$ turns into the multiplication operator by the phase function,
\[
\mathcal{F}S_j\mathcal{F}^{-1}\widehat{\psi}(\vartheta)=e^{i\theta_j}\widehat{\psi}(\vartheta),\qquad \widehat{\psi}\in L^2(\TT^d)\cong L^2(\widehat{\Gamma}).
\]

In a \emph{tight-binding model} with \emph{$m$ bands}, we would have $\mathcal{F}:\ell^2(\Gamma;\CC^m)\cong L^2(\widehat{\Gamma};\CC^m)$ instead. A \emph{finite-range Hamiltonian} on $\ell^2(\Gamma;\CC^m)$ is a self-adjoint operator $H=H^*$ expressible as an $m\times m$ matrix of polynomials in the translation operators $S_j, S_j^*$, with respect to any lattice basis. The Fourier transformed Hamiltonian $\mathcal{F}H\mathcal{F}^{-1}$ is then the multiplication operator on $L^2(\widehat{\Gamma};\CC^m)$ by some $m\times m$ Hermitian matrix-valued map on $\widehat{\Gamma}$, with entries being polynomial in $e^{\pm i\theta_j}$. It is customary to write this map as
\[
\chi\mapsto H(\chi)\in{\rm Herm}(m),\qquad \chi\in\widehat{\Gamma}.
\]
and refer to $H(\chi)$ as the \emph{Bloch Hamiltonian} at quasimomentum $\chi$. More generally, \emph{local} Hamiltonians are only required to be approximately finite-range in the sense that $\chi\mapsto H(\chi)$ is a \emph{smooth} assignment. This locality condition will be implicitly assumed throughout, and we will proceed to study $H$ through the \emph{differential topology} of $\widehat{\Gamma}$.

\medskip
Usually, $\Gamma$ is already given as $\ZZ^d$, and the ordered lattice basis gives an orientation on $\widehat{\Gamma}$. However, to get a Riemannian metric on $\widehat{\Gamma}$, we would need further geometric data of how $\Gamma$ sits inside a Euclidean space $V$ with inner product, see Section \ref{sec:geometry}. 

\section{Differential topology of band crossings}

Including multiplicity, each $H(\chi)\in{\rm Herm}(m)$ has $m$ real eigenvalues,
\[
\lambda_1(\chi)\leq\ldots\leq \lambda_m(\chi). 
\]
Even though $H$ is a smooth map, the eigenvalue functions, or energy \emph{bands}, $\lambda_i\equiv \lambda_i(\chi)$ may not be smooth, with issues arising precisely when degeneracies occur. It was already observed in \cite{vNW} that eigenvalue degeneracy generically occurs on \emph{codimension-3} submanifolds. \emph{From now on, we will work in $d=3$, so $\widehat{\Gamma}$ is a 3-torus.}

\subsection{Local topology of band crossings}\label{sec:local.crossings}
Let us assume that over some contractible open subset $\Omega\subset \widehat{\Gamma}$, we have the lowest two eigenvalues of $H(\chi)$ being isolated from the others,
\begin{equation}
\lambda_1(\chi)\leq \lambda_2(\chi)<\lambda_3(\chi)\leq \ldots,\qquad \chi\in\Omega.\label{eqn:isolated.from.rest}
\end{equation}
So over $\Omega$, we have smoothly-varying 2-dimensional subspaces $\mathcal{S}_\chi\subset\CC^m$, each given by the sum of the lowest two eigenspaces. Let us consider the Bloch Hamiltonians $H(\chi)$ truncated to $\mathcal{S}_\chi$. These are Hermitian endomorphisms, which we denote with the same symbol $H(\chi)$.

Since $\Omega$ is contractible, we can pick some smooth orthonormal frame over $\Omega$, so all the $\mathcal{S}_\chi$ become identified with a common reference $\CC^2$. Then the endomorphisms $H(\chi)$ become $2\times 2$ Hermitian matrices. A standard choice of basis for ${\rm Herm}(2)$ is
\begin{equation}
\mathbf{1}=\begin{pmatrix}1 & 0 \\ 0 & 1\end{pmatrix},\;\; \sigma_1=\begin{pmatrix} 0 & 1 \\ 1 & 0 \end{pmatrix},\;\; \sigma_2=\begin{pmatrix} 0 & -i\\ i & 0 \end{pmatrix},\;\; \sigma_3=\begin{pmatrix} 1 & 0 \\ 0 & -1\end{pmatrix},\label{eqn:Pauli.basis}
\end{equation}
with $\{\sigma_i\}_{i=1,2,3}$, spanning the \emph{traceless} Hermitian matrices ${\rm Herm}_0(2)$. Then we obtain a parametrization
\begin{equation}
H(\chi)=h_0(\chi)\mathbf{1}+\sum_{i=1}^3 h_i(\chi)\sigma_i=:h_0(\chi)\mathbf{1}+\mathbf{h}(\chi)\cdot\mathbf{\sigma},\label{eqn:generic.2x2}
\end{equation}
by some smooth functions
\[
h_0:\Omega\to\RR,\qquad\mathbf{h}\equiv (h_1,h_2,h_3):\Omega\to\RR^3.
\]
The traceless part of $H(\chi)$ squares to a scalar matrix,
\[
(H-h_0\mathbf{1})^2(\chi)=(\mathbf{h}(\chi)\cdot\mathbf{\sigma})^2=|\mathbf{h}(\chi)|^2 \mathbf{1}.
\]
It follows that the eigenvalue functions for $H(\chi)$ are
\[
\lambda_1(\chi)=h_0(\chi)-|\mathbf{h}(\chi)|,\qquad \lambda_2(\chi)=h_0(\chi)+|\mathbf{h}(\chi)|,\qquad \chi\in\Omega,
\]
which degenerate exactly when $\chi$ satisfies $\mathbf{h}(\chi)=(0,0,0)$.

\subsubsection{Local topological index of band crossings}\label{sec:local.topology.crossings}
As $H(\cdot)$ is a 3-parameter family, eigenvalue crossings generically occur at a set of points. More precisely, the vector field $\mathbf{h}$ over $\Omega\subset\widehat{\Gamma}$ generically and transversally intersects the zero vector field at a discrete set of points \cite{GP,Milnor}, called the \emph{zeroes} of $\mathbf{h}$.

By definition, a \emph{Weyl point} $w\in\Omega$ is an isolated zero of $\mathbf{h}$. Choose a small ball $B^3_w\subset\Omega$ centered at $w$ and containing no other Weyl point, then normalization of $\mathbf{h}$ makes sense over $B^3_w\setminus\{w\}$. In particular, we have a map between 2-spheres,
\[
\tfrac{\mathbf{h}}{|\mathbf{h}|}:S^2_w=\partial B^3_w\to S^2\subset \RR^3.
\]
By definition, the \emph{local index}, or \emph{local charge}, of $\mathbf{h}$ at the Weyl point $w$ is the degree of this map,
\begin{equation}
{\rm Ind}_w(\mathbf{h}):=\deg\Big(\tfrac{\mathbf{h}}{|\mathbf{h}|}:S^2_w\to S^2\Big)\in\ZZ\label{eqn:local.index}
\end{equation}
Here, we recall that the \emph{degree} of a smooth map $f:M_1\to M_2$ between compact oriented manifolds of the same dimension (with $M_2$ connected) is the generic number of points in the preimage $f^{-1}\{v\}$, $v\in M_2$, counted with signs, and it is a \emph{homotopy invariant}; see \S3.3 of \cite{GP} for a pedagogical treatment. 

We say that $w$ is a \emph{non-degenerate} Weyl point if the derivative $(d\mathbf{h})_w:\RR^3\to\RR^3$ at $w$ is invertible. In local coordinates, say $(k_1, k_2, k_3)$, for $\Omega$, we have
\[
(d\mathbf{h})_w=\begin{pmatrix} \frac{\partial h_i}{\partial k_j}\Big|_w\end{pmatrix}_{i,j=1,2,3}
\]
being the Jacobian matrix of first partial derivatives at $w$. It is a standard result \cite{GP,Milnor} that for non-degenerate zeroes,
\begin{equation}
{\rm Ind}_w(\mathbf{h})={\rm sgn}\det((d\mathbf{h})_w)=\pm 1.\label{eqn:non.degenerate.index}
\end{equation}
Furthermore, if $w$ is \emph{degenerate}, a generic perturbation of $\mathbf{h}$ near $w$ will split $w$ into a collection of non-degenerate Weyl points, whose index sum equals the original index at $w$. 
So Eq.\ \eqref{eqn:non.degenerate.index} is generically applicable.

To understand the significance of ${\rm Ind}_w(\mathbf{h})$, suppose $\mathbf{h}:B^3_w\to\RR^3$ can be smoothly deformed to a \emph{nowhere vanishing} map $\mathbf{h}^\prime$. So there is no longer any Weyl point in $B^3_w$, and
\[
\tfrac{\mathbf{h}^\prime}{|\mathbf{h}^\prime|}:B^3_w\to \RR^3\setminus\{0\}\to S^2
\]
makes sense everywhere in $B^3_w$. Necessarily, the degree of $\tfrac{\mathbf{h}^\prime}{|\mathbf{h}^\prime|}:S^2_w\to S^2$ vanishes, i.e., ${\rm Ind}_w(\mathbf{h}^\prime)=0$. But we would also have a deformation $\tfrac{\mathbf{h}}{|\mathbf{h}|}\sim \tfrac{\mathbf{h}^\prime}{|\mathbf{h}^\prime|}$ of maps $S^2_w\to S^2$. By homotopy invariance of degree, ${\rm Ind}_w(\mathbf{h})=0$ as well. 

In summary: \emph{A non-vanishing local index of $\mathbf{h}$ at $w$ topologically protects the eigenvalue crossing at $w$ from acquiring a spectral gap via deformations of $\mathbf{h}$  near $w$} (and away from other eigenvalue crossings).

\begin{example}\label{ex:minimal.example}
A minimal 2-band toy model, written in lattice-adapted coordinates $\chi\leftrightarrow\vartheta=(\theta_1,\theta_2,\theta_3)$, taken from \cite{AMV}, is $H(\vartheta)=\mathbf{h}(\vartheta)\cdot\sigma$ with
\[
h_1(\vartheta)=\sin \theta_1,\;\; h_2(\vartheta)=\sin \theta_2,\;\; h_3(\vartheta)=2+t-\sum_{i=1}^3\cos \theta_i,
\]
where $t\in\RR$ is a parameter. For $|t|<1$, non-degenerate Weyl points occur at $\vartheta=(0,0,\pm \cos^{-1}t)$, with $\pm 1$ local indices. As $t$ is increased, the Weyl points are created when $t=-1$ at $\vartheta=(0,0,\pi)$, then they are annihilated when $t=1$ at $\vartheta=(0,0,0)$.
\end{example}

\subsection{Global invariant description}\label{sec:invariant.description}
It is important to recognize that the parametrization of $H$, Eq.\ \eqref{eqn:generic.2x2}, has an ambiguity due to the ${\rm U}(2)$ gauge freedom in identifying $\mathcal{S}_\chi\cong\CC^2$. Nevertheless, it is easily verified that a gauge transformation $\chi\mapsto U(\chi)\in{\rm U}(2)$ causes the traceless part of $H(\chi)$ to be conjugated to
\begin{equation}
U(\chi)\big(\mathbf{h}(\chi)\cdot\mathbf{\sigma}\big)U^{-1}(\chi)=\mathbf{h}^\prime(\chi)\cdot\mathbf{\sigma},\label{eqn:spin.rotation}
\end{equation}
with the transformed 3-vectors $\mathbf{h}^\prime(\chi)$ being orthogonally rotated from $\mathbf{h}(\chi)$,
\begin{equation}
\mathbf{h}^\prime(\chi)=R(\chi)\cdot\mathbf{h}(\chi),\qquad R(\chi)\in{\rm SO}(3).\label{eqn:orthogonal.rotation}
\end{equation}
For $U(\chi)\in{\rm SU}(2)\cong{\rm Spin}(3)$, this is the familiar physicist derivation of the spin double-cover of ${\rm SO}(3)$. More generally, we have ${\rm U}(2)\cong {\rm Spin}^{\rm c}(3)$, and Eq.\ \eqref{eqn:spin.rotation}-\eqref{eqn:orthogonal.rotation} realizes ${\rm PU}(2)\cong {\rm SO}(3)$ for the \emph{projective} unitary group. The $\mathcal{S}_\chi$ are sometimes called ``spinor spaces'', on which unitaries act as ``spin-rotations''. Note that a field of ${\rm SO}(3)$-rotations over $S^2_w$ can be homotoped to the identity, so the above gauge freedom does not affect the topological indices ${\rm Ind}_w(\mathbf{h})$.

Suppose the lowest two eigenvalues of $H(\chi)$ are isolated from the others, for all points $\chi\in\widehat{\Gamma}$. So $\mathcal{S}_\chi\subset\CC^m$ is well-defined for all $\chi\in\widehat{\Gamma}$. This is the situation of a (possibly indirect) gap above the second energy band. We may then pass to the \emph{effective} two-band model given by the family of truncated endomorphisms $H(\chi)$ acting on $\mathcal{S}_\chi, \chi\in\widehat{\Gamma}$.

Invariantly, the family of subspaces $\mathcal{S}_\chi\subset\CC^m$ form a rank-2 Hermitian vector bundle $\mathcal{S}\to\widehat{\Gamma}$. Its corresponding bundle of Hermitian endomorphisms is a \emph{real} vector bundle ${\rm Herm}(\mathcal{S})\to \widehat{\Gamma}$ of rank $4$, and the effective family $\chi\mapsto H(\chi)$ is a \emph{global section} of ${\rm Herm}(\mathcal{S})$.

On ${\rm Herm}(\mathcal{S})$, there is a ${\rm U}(2)$-invariant fibrewise inner product $\langle H_1,H_2\rangle=\tfrac{1}{2}{\rm Tr}(H_1H_2)$. The corresponding norm is denoted $||\cdot||$. The ``trace-ful'' part of ${\rm Herm}(\mathcal{S})$ is a trivial rank-1 subbundle $\underline{\RR}$, while the traceless part ${\rm Herm}_0(\mathcal{S})$ is a subbundle of rank 3. They are orthogonal to each other. Due to its action by conjugation, the structure group on ${\rm Herm}_0(\mathcal{S})$ is effectively ${\rm PU}(2)\cong{\rm SO}(3)$.

The traceless Bloch Hamiltonians, $H-h_0\mathbf{1}$, thus constitute a \emph{global section} of ${\rm Herm}_0(\mathcal{S})$. Over a local patch $\Omega\subset\widehat{\Gamma}$, we may choose some trivialization $\mathcal{S}|_{\Omega}\cong\Omega\times \CC^2$, and identify ${\rm Herm}_0(\mathcal{S})|_{\Omega}\cong\Omega\times\RR^3$ using the local expansion $H-h_0\mathbf{1}\sim\mathbf{h}\cdot\sigma$, as in Section \ref{sec:local.crossings}. The eigenvalues of $H(\chi)$ are $h_0(\chi)\pm ||H(\chi)-h_0(\chi)\mathbf{1}||$, so the spectral gap of $H(\chi)$ closes precisely at the zeroes of the traceless section \mbox{$H-h_0\mathbf{1}$}.

Regarding the fibre orientation on ${\rm Herm}_0(\mathcal{S})$, this comes from the choice of ordered basis for ${\rm Herm}_0(2)$ in Eq.\ \eqref{eqn:Pauli.basis}, which satisfies $-i\sigma_1\sigma_2\sigma_3=\mathbf{1}$; see Section \ref{sec:relativistic.Dirac}.

\subsection{Global cancellation of local indices}
The coordinate-free understanding of topological protection of local band crossings pays dividends when trying to understand its \emph{global} aspects.

\subsubsection{Degree formula and Stokes' theorem}
A globally trivializability condition, $\mathcal{S}\cong \widehat{\Gamma}\times\CC^2$, is typically assumed on the lowest two bands, i.e., the effective model in question. In this case, ${\rm Herm}_0(\mathcal{S})\cong \widehat{\Gamma}\times\RR^3$ is likewise globally trivializable. As explained in Section \ref{sec:local.topology.crossings}, we have $H$ describable by a smooth assignment of matrices,
\[
H=h_0\mathbf{1}+\mathbf{h}\cdot\sigma:\widehat{\Gamma}\to {\rm Herm}(2),
\]
where $\mathbf{h}:\widehat{\Gamma}\to\RR^3$ is a globally defined 3-component function, whose vanishing set $W\subset\widehat{\Gamma}$ is the set of Weyl points. Write $B^3_W:=\sqcup_{w\in W} B^3_w$ for the small neighbourhood of the Weyl points. Note that $\widehat{\Gamma}\setminus B^3_W$ is a compact manifold with boundary being the union of the small spheres $S^2_w$ enclosing the Weyl points. By construction, we have a smooth normalized map
\[
\tfrac{\mathbf{h}}{|\mathbf{h}|}:\widehat{\Gamma}\setminus B^3_W\to S^2,
\]
whose restriction to each $S^2_w$ has degree equal to the local index at $w$ (Eq.\ \eqref{eqn:local.index}).

Now, on $S^2$ with angular coordinates $(\theta,\varphi)$, consider the closed $2$-form $\eta=\sin\theta\,d\theta\wedge d\varphi$ with $\int_{S^2}\eta=4\pi$. The pullback $\left(\tfrac{\mathbf{h}}{|\mathbf{h}|}\right)^*\eta$ is a closed $2$-form on $\widehat{\Gamma}\setminus W$, and an application of Stokes' theorem gives
\begin{equation}
-\int_{\widehat{\Gamma}\setminus B^3_W} \underbrace{d\left(\tfrac{\mathbf{h}}{|\mathbf{h}|}\right)^*\eta}_{0} =\sum_{w\in W}\int_{S^2_w}\left(\tfrac{\mathbf{h}}{|\mathbf{h}|}\right)^*\eta.\label{eqn:Stokes}
\end{equation}
The \emph{degree formula} (\S 8 of \cite{GP}) reads
\[
\int_{S^2_w}\left(\tfrac{\mathbf{h}}{|\mathbf{h}|}\right)^*\eta={\rm deg}\Big(\tfrac{\mathbf{h}}{|\mathbf{h}|}:S^2_w\to S^2\Big)\cdot \int_{S^2}\eta,
\]
so that Eq.\ \eqref{eqn:Stokes} is simply the \emph{global charge-cancellation} constraint,
\begin{equation*}
0=\sum_{w\in W}{\rm deg}\Big(\tfrac{\mathbf{h}}{|\mathbf{h}|}:S^2_w\to S^2\Big)\equiv \sum_{w\in W}{\rm Ind}_w(\mathbf{h}).
\end{equation*}
This vanishing result implies that if there is a single pair of Weyl points, one of which has local index $n\neq 0$, then the other one must have local index $-n$. 

In the literature, an analogous result of Nielsen--Ninomiya for chiral lattice fermions \cite{NN} is often quoted, but the precise setting there is somewhat different. The simplified direct argument above follows \S 1.3 of \cite{Witten}.

\subsubsection{Poincar\'{e}--Hopf index theorem}
As explained in Section \ref{sec:invariant.description}, in general, ${\rm Herm}_0(\mathcal{S})\to\widehat{\Gamma}$ is an oriented rank-3 vector bundle over the oriented 3-torus $\widehat{\Gamma}$, and it may not be trivializable.
It has an \emph{Euler class} $e({\rm Herm}_0(\mathcal{S}))$ in the cohomology group $H^3(\widehat{\Gamma};\ZZ)$ (see \S9 of \cite{MS}, \S VIII.1 of \cite{GHV}).

Recall that $H-h_0\mathbf{1}$ is a global section of ${\rm Herm}_0(\mathcal{S})$, vanishing at the Weyl point set $W$. The local indices of $H-h_0\mathbf{1}$ are defined as before, by using a local trivialization. The global constraint on these local indices is the celebrated 
\emph{Euler--Poincar\'{e}--Hopf} theorem (e.g., \S VIII.4 Theorem II \cite{GHV})
\begin{equation}
[\widehat{\Gamma}]\cdot\sum_{w\in W}{\rm Ind}_w(H-h_0\mathbf{1})=e({\rm Herm}_0(\mathcal{S})),\label{eqn:PH.cancellation}
\end{equation}
where $[\widehat{\Gamma}]$ denotes the orientation class generating $H^3(\widehat{\Gamma};\ZZ)\cong\ZZ$. In the special case of tangent vector bundles, the fact that the index sum equals the Euler characteristic of the base manifold, is colloquially known as the ``hairy-ball theorem'' (see \cite{GP,Milnor} for a pedagogical treatment).

The key point is that ${\rm Herm}_0(\mathcal{S})$ has odd fibre dimension, which implies that its Euler class is minus of itself (Property 9.4 of \cite{MS}), so it must vanish by virtue of $H^3(\widehat{\Gamma};\ZZ)\cong \ZZ$. Therefore, we obtain a more general charge-cancellation condition,
\begin{equation}
\sum_{w\in W}{\rm Ind}_w(H-h_0\mathbf{1})=0,\label{eqn:general.charge.cancellation}
\end{equation}
whether or not $\mathcal{S}$ is trivializable.

\section{Global topological invariant of Weyl semimetals}

\subsection{Global Chern class of Weyl semimetals}
The non-trivializability of a complex line bundle $\mathcal{L}\to X$ is characterized by its global Chern class
\[
c_1(\mathcal{L})\in H^2(X;\ZZ).
\]
Here, $H^2(X;\ZZ)$ is a certain abelian group, called an \emph{integral cohomology class}, whose elements may be represented by closed differential 2-forms on $X$ that integrate to integers on closed 2-submanifolds of $X$, modulo exact 2-forms (see \cite{BT}, Appendix C of \cite{MS}). The integers thus obtained are called the \emph{Chern numbers} of $\mathcal{L}$ (over the 2-submanifold in question). An important implication of $c_1(\mathcal{L})\neq 0$ is the non-existence of a smooth nowhere-zero assignment $\chi\mapsto v(\chi)\in\mathcal{L}_\chi$ globally over $X$.

For example, away from the Weyl point set, there is a well-defined \emph{Weyl semimetal line bundle} $\mathcal{L}_1\to\widehat{\Gamma}\setminus W$ comprising the lowest-energy eigenspaces, \mbox{$(\mathcal{L}_1)_\chi\subset\CC^m$}. Actually, for the purposes of this section, we do not even need to assume isolation of the lowest two energy bands, Eq.\ \eqref{eqn:isolated.from.rest}.

\subsection{Local Chern classes of Weyl semimetals}
By restricting $\mathcal{L}_1$ to $S^2_w$, we obtain the \emph{local Chern classes}
\[
c_1(\mathcal{L}_1|_{S^2_w})\in H^2(S^2_w;\ZZ),\qquad w\in W.
\]
Although unnecessary for what follows, we may understand these local Chern classes as local indices of vector fields as follows. 

Over some small 3-ball $B^3_w\ni w$, we can trivialize the lowest two bands, and express the effective 2-band Bloch Hamiltonian as $H(\chi)=h_0\mathbf{1}+\mathbf{h}\cdot\sigma$ for a smooth map $\mathbf{h}:B^3_w\to \RR^3$ vanishing only at $w$ (Eq.\ \eqref{eqn:generic.2x2} of Section \ref{sec:local.crossings}). 

For each $\chi\in S^2_w$, the lowest-energy eigenspace $(\mathcal{L}_1)_{\chi}$ is a certain complex line in $\CC^2$ (i.e., an element of $\CC\PP^1$); namely, the $-1$ eigenspace of the ``spin operator'' $\tfrac{\mathbf{h}(\chi)}{|\mathbf{h}(\chi)|}\cdot\sigma$. As explained in Section \ref{sec:Bloch.tautological}, $\mathcal{L}_1|_{\chi}$ is exactly the orthogonal complement of the complex line that tautologically corresponds to the point $\tfrac{\mathbf{h}(\chi)}{|\mathbf{h}(\chi)|}\in S^2\cong \CC\PP^1$. In other words, the local line bundle $\mathcal{L}_1\to S^2_w$ is the pullback 
\[
\mathcal{L}_1|_{S^2_w}\cong\left(\tfrac{\mathbf{h}}{|\mathbf{h}|}\big|_{S^2_w}\right)^*\mathcal{L}_{\rm taut}^\perp.
\]
The corresponding local Chern class is
\begin{align*}
c_1(\mathcal{L}_1|_{S^2_w})&=c_1\left(\left(\tfrac{\mathbf{h}}{|\mathbf{h}|}\big|_{S^2_w}\right)^*\mathcal{L}_{\rm taut}^\perp\right)\\
&=\left(\tfrac{\mathbf{h}}{|\mathbf{h}|}\big|_{S^2_w}\right)^*c_1(\mathcal{L}_{\rm taut}^\perp)\in H^2(S^2_w;\ZZ).
\end{align*}
Above, we used the naturality property of $c_1$ to commute it with the pullback operation. The pullback map on cohomology groups,
\[
\left(\tfrac{\mathbf{h}}{|\mathbf{h}|}|_{S^2_w}\right)^*:\underbrace{H^2(S^2;\ZZ)}_{\ZZ}\to \underbrace{H^2(S^2_w;\ZZ)}_{\ZZ},
\]
depends only on the homotopy class of \mbox{$\tfrac{\mathbf{h}}{|\mathbf{h}|}\big|_{S^2_w}:S^2_w\to S^2$}, which is exactly its degree by Hopf's degree theorem \cite{GP,Milnor}. But this degree is precisely the local index at $w$, by Eq.\ \eqref{eqn:local.index}. Because $c_1(\mathcal{L}_{\rm taut}^\perp)=1$ (see Section \ref{sec:Bloch.tautological}), this means that
\begin{equation*}
c_1(\mathcal{L}_1|_{S^2_w})={\rm Ind}_w(\mathbf{h}).
%\label{eqn:local.Chern.equals.local.index}
\end{equation*}

\subsection{Mayer--Vietoris sequence}
A natural question is whether the global Weyl semimetal invariant $c_1(\mathcal{L}_1)$ is already completely determined by the local Chern classes $c_1(\mathcal{L}_1|_{S^2_w})$, and the answer is \emph{no}. For instance, the ``global connectivity'' of the Weyl points inside $\widehat{\Gamma}$ cannot be captured by the local Chern classes. 

Because Chern classes are cohomology groups of the base space, they obey a locality principle with respect to decompositions of the base space. This is called the \emph{Mayer--Vietoris} (MV) sequence \cite{BT}.

In the Weyl semimetal context, the relevant decomposition of the Brillouin torus is \cite{MT2, MT}
\[
\widehat{\Gamma}=(\widehat{\Gamma}\setminus W)\cup B^3_W.
\]
The overlap region of the above covering can be retracted to the small 2-spheres around the Weyl points,
\[
(\widehat{\Gamma}\setminus W) \cap B^3_W\sim_{\rm retract} S^2_W:=\bigsqcup_{w\in W} S^2_w.
\]
Intuitively, a cohomology class over $\widehat{\Gamma}$ should split into a contribution from \mbox{$\widehat{\Gamma}\setminus W$} plus a contribution from $B^3_W$, minus the contribution from the overlap $S^2_W$. This would be true if we were just considering representatives (e.g.\ closed differential forms), but at the cohomology class level (i.e.\ modulo exact forms), there are lower/higher degree corrections. The MV-sequence systematically accounts for this, and it reads
\begin{equation}
\begin{tikzcd}
 & \qquad\cdots\qquad \rar 
             \ar[draw=none]{d}[name=X, anchor=center]{}
    & H^1(S^2_W) \ar[rounded corners,
            to path={ -- ([xshift=2ex]\tikztostart.east)
                      |- (X.center) \tikztonodes
                      -| ([xshift=-2ex]\tikztotarget.west)
                      -- (\tikztotarget)}]{dll}[at end]{\delta} 
      \\ 
H^2(\widehat{\Gamma}) \rar & \begin{array}{c}H^2(\widehat{\Gamma}\setminus W) \\ \oplus\; H^2(B^3_W)\end{array} \rar
             \ar[draw=none]{d}[name=X, anchor=center]{}
    & H^2(S^2_W) \ar[rounded corners,
            to path={ -- ([xshift=2ex]\tikztostart.east)
                      |- (X.center) \tikztonodes
                      -| ([xshift=-2ex]\tikztotarget.west)
                      -- (\tikztotarget)}]{dll}[at end]{\delta} 
       \\      
 H^3(\widehat{\Gamma}) \rar &  \begin{array}{c}H^3(\widehat{\Gamma}\setminus W)\\ \oplus\; H^3(B^3_W)\end{array} \rar &  \quad\cdots,
\end{tikzcd}\label{eqn:cohom.MV.longer}
\end{equation}
where we suppress the $\ZZ$-coefficients in $H^\bullet(\cdot)$. The $\delta$ symbol indicates a \emph{connecting homomorphism}, while the unmarked arrows are simply restriction maps. 
For non-empty $W$, basic calculations in algebraic topology give:
\begin{itemize}
\item $H^2(B^3_W)=0=H^3(B^3_W)$, $H^1(S^2_W)=0$, and \mbox{$H^3(\widehat{\Gamma}\setminus W)=0$}.
\item $H^2(S^2_W)=\bigoplus_{w\in W} H^2(S^2_w)=\bigoplus_{w\in W}\ZZ$, the group of ``local charges''.
\item $H^2(\widehat{\Gamma})\cong\ZZ^3$ and $H^3(\widehat{\Gamma})\cong\ZZ$.
\end{itemize} 
Then the MV-sequence \eqref{eqn:cohom.MV.longer} shortens to
\begin{equation}
0\to \underbrace{H^2(\widehat{\Gamma})}_{\ZZ^3}\to H^2(\widehat{\Gamma}\setminus W)\to \underbrace{H^2(S^2_W)}_{\ZZ^{|W|}}\overset{\delta}{\to} \underbrace{H^3(\widehat{\Gamma})}_{\ZZ}\to 0\label{eqn:cohom.MV},
\end{equation}
where $\delta:\ZZ^{|W|}\to\ZZ$ is the summation map (see \cite{MT} for details). 

Importantly, Eq.\ \eqref{eqn:cohom.MV.longer}--\eqref{eqn:cohom.MV} are \emph{exact} sequences of abelian groups, meaning that the kernel of each homomorphism exactly equals the range of the previous homomorphism. This exactness has the following consequences:
\begin{itemize}
\item A list of integers $(n_w)_{w\in W}\in \ZZ^{|W|}$ is consistent with the local charge data of a Weyl semimetal line bundle $\mathcal{L}_1\to\widehat{\Gamma}\setminus W$ iff charge-cancellation holds, $\sum_{w\in W}n_w=0$.
\item A Weyl semimetal line bundle extends to all of $\widehat{\Gamma}$ (i.e., it becomes insulating) iff all its local charges are zero. If $W$ only has one point, its local charge must vanish, and extendibility would be automatic.
\item Given local charges $(n_w)_{w\in W}\in \ZZ^{|W|}$ summing to zero, the global Chern class of a Weyl semimetal with these local charges has an indeterminacy group $H^2(\widehat{\Gamma})$.
\end{itemize}
The second consequence is more subtle, as illustrated by the example below.

\begin{figure}
\centering
\begin{tikzpicture}
\draw (0,4)--(0,0)--(4,0)--(4,4)--(0,4)--(1,5)--(5,5)--(5,1)--(4,0);
\draw (4,4)--(5,5);
\draw[dashed] (0,0)--(1,1)--(5,1);
\draw[dashed] (1,1)--(1,5);
\node at (2.1,2.5) {$w_+$};
\node at (3.5,3) {$w_-$};
\node at (1.4,3.7) {$T_L$};
\node at (2.8,3.7) {$T_R$};
\node at (1.8,0.7) {$\ell_L$};
\node at (3,0.7) {$\ell_R$};
\node at (2.2,0.3) {$\bullet$};
\node at (3.5,0.5) {$\bullet$};
\draw (2.1,2.5) circle (0.4cm);
\draw (3.5,3) circle (0.4cm);
\draw[thin] (1.1,0)--(1.1,4)--(2.1,5);
\draw[thin, dashed] (2.1,5)--(2.1,1)--(1.1,0);
\draw[thin] (2.3,0)--(2.3,4)--(3.3,5);
\draw[thin, dashed] (3.3,5)--(3.3,1)--(2.3,0);
\draw[dotted] (2.2,2.3)--(2.2,0.3);
\draw[dotted] (3.5,2.8)--(3.5,0.5);
\draw[ultra thick,dotted] (2.3,2.5)--(3.2,3);
\draw[ultra thick,dashed] (3.8,3)--(4.3,2.9);
\draw[ultra thick,dashed] (0.3,2.9)--(1.9,2.5);
\draw[thick] (2.2,0.3) to [bend right=30] (3.5,0.5);
\end{tikzpicture}
\caption{Brillouin 3-torus represented as a cube, with opposite faces identified. The ``planes'' $T_L, T_R$ are 2-tori ``enclosing'' $w_+$. They project onto loops $\ell_L, \ell_R$ on the surface Brillouin 2-torus. Two inequivalent Dirac strings (thick dotted line and thick dashed line), both consistent with $w_\pm$ having local index $\pm 1$, are drawn. The correct Dirac string for a Weyl semimetal Hamiltonian projects onto its resultant surface Fermi arc (thick curve).}\label{fig:two.Weyl.points}
\end{figure}
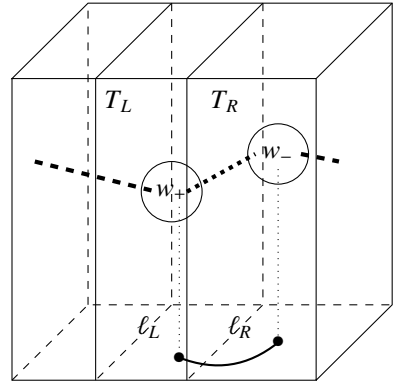

\begin{example}
Suppose there are only two Weyl points, $w_+, w_-\in\widehat{\Gamma}$, with respective local charges $+1, -1$, as illustrated in Figure \ref{fig:two.Weyl.points}. The semimetal line bundle $\mathcal{L}_1$ has a well-defined $\mathfrak{u}(1)$-valued (closed) curvature 2-form $\mathcal{F}$ on $\widehat{\Gamma}\setminus\{w_+,w_-\}$. Consider a pair of 2-tori, $T_L, T_R$, ``enclosing'' $w_+$. We can integrate $d\mathcal{F}=0$ over the region $\mathcal{V}$ of $\widehat{\Gamma}$ bounded by $T_L, T_R$ and $S^2_{w_+}$. By Stokes' theorem, this is
\[
0=\frac{i}{2\pi}\int_\mathcal{V}d\mathcal{F}=\frac{i}{2\pi}\int_{\partial\mathcal{V}=-T_L+T_R-S^2_{w_+}}\mathcal{F}=-n_L+n_R-1,
\]
where $n_L, n_R$ are the Chern numbers obtained by integrating $i\mathcal{F}/2\pi$ over $T_L$ and $T_R$ respectively. Thus, the local index $+1$ of $w_+$ only determines \mbox{$n_R-n_L=1$}, but not $n_L$ or $n_R$ individually. We could have, e.g., $n_R=1$ (``non-trivial in between $w_+$ and $w_-$'') or $n_L=-1$ (``trivial in between $w_+$ and $w_-$''). Repeating this construction for 2-tori parallel to the other two independent choices of direction-pairs, we deduce the $\ZZ^3\cong H^2(\widehat{\Gamma})$ indeterminacy in the global Chern class. 
\end{example}

\section{Dual Dirac string description}
\subsection{Magnetic monopoles and Dirac strings}\label{sec:magnetic.monopole}
A convenient way to represent the global Chern class of a Weyl semimetal is to Poincar\'{e} dualize from cohomology to homology, as explained in \cite{MT2,MT}. In physics language, one passes from Chern classes to Dirac strings. 

Let us illustrate this from the well-known case of a single Dirac magnetic monopole of charge $g/2$ at the origin of $\RR^3$. In spherical coordinates $(r,\theta,\varphi)$ with orthonormal frame $\{\hat{r},\hat\theta,\hat{\varphi}\}$, a candidate vector potential is
\[
\mathbf{A}=g\frac{1-\cos\theta}{2r\sin\theta}\hat{\varphi},\qquad \mathbf{B}=\nabla\times\mathbf{A}=\frac{g}{2r^2}\hat{r}
\]
The radial monopole field $\mathbf{B}$ is well-defined on $\RR^3\setminus \{0\}$, and has flux $2\pi g$ through any 2-sphere enclosing the origin. But $\mathbf{A}$ is ill-defined on the ``Dirac string'' $\theta=\pi$.

In modern language, $\mathbf{B}$ is regarded, via Hodge duality, as the differential 2-form \mbox{$B=\tfrac{g}{2}\sin\theta\,d\theta\wedge d\varphi$} over $\RR^3\setminus\{0\}$. Despite $B$ being closed, $dB=0$, we cannot write $B=dA$ for any 1-form defined over all of $\RR^3\setminus\{0\}$. The best we can do is to choose, e.g., $A=g\sin^2\tfrac{\theta}{2}\,d\varphi$, which blows up along $\theta=\pi$. Alternatively, $A^\prime=-g \cos^2\tfrac{\theta}{2}\,d\varphi$ also works, but blows up along $\theta=0$. Either way, the failure occurs on a Dirac string connecting the origin to infinity. Regarding the ambiguity of whether to use $A$ or $A^\prime$ on their overlapping region of validity (i.e.\ away from the $z$-axis), this is resolved by recognising that $\mathcal{A}=-iA$ and $\mathcal{A}^\prime=-iA^\prime$ are merely gauge-dependent representations of a connection on a ${\rm U}(1)$-line bundle over $\RR^3\setminus\{0\}$. (In physics, it is usual to absorb a factor of $i$ in these expressions.) These choices are related by the gauge transformation $U(r,\theta,\varphi)=\exp(ig\varphi)$, which is well-defined away from the $z$-axis, provided Dirac's quantization condition for the monopole charge, $g\in \mathbb{Z}$, holds. The curvature of the connection is the globally defined 2-form
\[
\mathcal{F}=d\mathcal{A}=d\mathcal{A}^\prime=-\frac{ig}{2}\sin\theta\,d\theta\wedge d\varphi,
\]
which integrates over any 2-sphere $S^2_0$ enclosing the origin to give the Chern number
\[
\frac{i}{2\pi}\int_{S^2_0}\mathcal{F}=g\in\ZZ.
\]
No particular geometric Dirac string is distinguished; what matters is that \emph{some} Dirac string must always be excluded from $\RR^3\setminus\{0\}$ when representing the connection as a 1-form $A$. 

Topologically speaking, $\RR^3\setminus\{0\}$ is the 3-sphere $S^3$ with two points removed (the origin and the point at infinity). Then Poincar\'{e}--Lefschetz duality is the isomorphism (e.g., \S VI.8 Theorem 8.3 of \cite{Bredon})
\[
\underbrace{H^2(S^3\setminus\{0,\infty\})}_{{\rm Chern\;class}}\cong H_1(S^3,\{0,\infty\}), 
\]
where the right side is the first relative homology group, whose classes are represented by 1-manifolds with ends on $0$ and $\infty$. In other words, Dirac strings provide a dual representation of the Chern class obstruction to trivializing (i.e., choosing a globally defined gauge for) the monopole line bundle.

Let us also mention that the Chern class obstruction is \emph{topological}, in the sense that it is independent of the choice of connection on the line bundle. The connection (and its curvature) is extra \emph{geometric} data.

\subsection{Tautological line bundle and Bloch sphere}\label{sec:Bloch.tautological}
Consider the 2-sphere $S^2\subset\RR^3$, parametrized by angular coordinates $(\theta,\varphi)$. The spin operator in direction $(\theta,\varphi)$ is
\[
(\sin\theta\cos\varphi, \sin\theta\sin\varphi, \cos\theta)\cdot\sigma=\begin{pmatrix} \cos\theta & \sin\theta e^{-i\varphi}\\ \sin\theta e^{i\varphi} & -\cos\theta\end{pmatrix}.
\]
Let us consider the line bundle $\mathcal{L}^+\to S^2$, whose fibre $\mathcal{L}^+_{(\theta,\varphi)}$ at $(\theta,\varphi)$ is the $+1$-eigenspace of the above spin operator, 
\[
\mathcal{L}^+_{(\theta,\varphi)}={\rm span}_\CC\binom{\cos\frac{\theta}{2}}{\sin\frac{\theta}{2}e^{i\varphi}}.
\]
(Note that this makes sense at $\theta=0,\pi$, despite the ill-definedness of $\varphi$ there.)
As is familiar from the physicists' \emph{Bloch sphere}, as $(\theta,\varphi)$ runs over all points of $S^2$, the $\mathcal{L}^+_{(\theta,\varphi)}$ run over the manifold of all 1-dimensional complex lines in $\CC^2$, i.e, $\CC\PP^1$.

So we have the identification
\[
S^2\ni (\theta,\varphi)\;\leftrightarrow\;{\rm span}_\CC\binom{\cos\frac{\theta}{2}}{\sin\frac{\theta}{2}e^{i\varphi}}\in\CC\PP^1,
\]
and the line $\mathcal{L}^+_{(\theta,\varphi)}$ is precisely the element of $\CC\PP^1$ that $(\theta,\varphi)$ corresponds to. In other words, $\mathcal{L}^+\to S^2$ is identified with the \emph{tautological line bundle} $\mathcal{L}_{\rm taut}\to\CC\PP^1$. Simiarly, the $(-1)$-eigenspace bundle is identified with the orthogonal complement line bundle $\mathcal{L}_{\rm taut}^\perp\to\CC\PP^1$.

The Chern class of $\mathcal{L}_{\rm taut}\to\CC\PP^1$ is, by a characterization of Chern classes \cite{MS}, a generator of $H^2(\CC\PP^1;\ZZ)\cong\ZZ$, which we choose to be $-1$. (See Example \ref{ex:tautological.connection} for a direct calculation.) Likewise, the Chern class of $\mathcal{L}_{\rm taut}^\perp\to\CC\PP^1$ is $+1$.

\subsection{Dirac string representation of Weyl semimetals}
Returning to Weyl semimetals, the same Poincar\'{e}--Lefschetz duality applies with $S^3$ replaced by $\widehat{\Gamma}$ and $\{0,\infty\}$ replaced by $W$,
\begin{equation}
H^2(\widehat{\Gamma}\setminus W)\cong H_1(\widehat{\Gamma},W).\label{eqn:co.homology.invariant}
\end{equation}
The right side of \eqref{eqn:co.homology.invariant} is represented by ``Dirac strings'' which are allowed to start/end on $W$ (closed loops with no endpoints are also allowed). The intersection number of the Dirac string with a closed 2-submanifold of \mbox{$\widehat{\Gamma}\setminus W$} gives the Chern number over that submanifold. As with the magnetic monopole, only the homology class of the Dirac string matters; particular representative strings are unphysical. 

For a Weyl semimetal, its (co)homology class in \eqref{eqn:co.homology.invariant} is preserved as long as no gap-closing occurs (other than at the Weyl points). The analysis is more subtle if we allow the Weyl points to move around inside $\widehat{\Gamma}$. For example, when $w_+$ and $w_-$ come together, a spectral gap may be opened (because the local charges now cancel). Conversely, a gap-opening procedure could equally occur elsewhere in $\widehat{\Gamma}$, and then the Weyl points $w_+, w_-$ may be moved back to their starting position. At the beginning and end of this gap-opening/closing procedure, the local charges are the same, yet the global topological invariant can change, if the Weyl point creation/annihilation process takes place over a non-contractible region in $\widehat{\Gamma}$, see Section 3 of \cite{MT}. Example \ref{ex:minimal.example} explicitly illustrates this principle, as explained in \cite{AMV} \S II.B.1.

\section{Differential geometry}\label{sec:geometry}
Geometry is relevant for the magnetic monopole (Section \ref{sec:magnetic.monopole}) in two ways. First, the Euclidean metric and orientation on \emph{position space} $\RR^3$ is used to convert between the radial magnetic field $\mathbf{B}$ and the 2-form $B$. Second, the connection on the monopole line bundle provides the notion of parallel transport in the line bundle, with spherically-symmetric curvature. The geometric situation for the Weyl semimetal is rather different. 

\subsection{Geometry of Brillouin torus}
Let $V$ be the group of translations of $d$-dimensional Euclidean (position) space. Note that the Euclidean space manifold has a Riemannian metric, and that $V$ is canonically identified with the tangent spaces of Euclidean space, thus $V$ is an inner product space. Similarly, momentum space $V^*$ may be viewed as the cotangent spaces, with the dual inner product. 

Given a lattice $\Gamma\subset V$, the Brillouin torus $\widehat{\Gamma}=V^*/\Gamma^\perp$ is now a Riemannian manifold. Note that $\Gamma$, therefore also $\Gamma^\perp$, is generally not cubic in the sense of admitting an orthonormal (or even orthogonal) lattice basis. Consequently, the Brillouin torus $\widehat{\Gamma}$ is generally not isometric to a ``standard'' $\TT\times\ldots\times\TT$ with product Riemannian metric. 

Now, if we have an orthonormal basis for $V$, thus for $V^*$, then momentum vectors $p\in V^*$ are labelled by $(p_1,\ldots,p_d)\in\RR^d$. Quasimomenta $\chi\in\widehat{\Gamma}$ are labelled by $d$-tuples $(k_1,\ldots,k_d)$, understood to be taken modulo $\Gamma^\perp\subset V^*\cong\RR^d$, so that they provide local coordinates for $\widehat{\Gamma}$.

\subsubsection{Dirac and Weyl operators}\label{sec:relativistic.Dirac}
In relativistic quantum mechanics and spin geometry, the Dirac/Weyl operator is a \emph{geometrically} canonical first-order differential operator \cite{LM} associated to the (semi-)Riemannian metric of physical space. For example, in Euclidean space $V\cong\RR^3$ with standard coordinates $(x_1,x_2,x_3)$, the bundle of Weyl spinors is $\RR^3\times\CC^2$, and the Weyl operators are
\[
H_\pm=\mp i\sum_{j=1}^3\partial_j\sigma_j,
\]
distinguished by how $\partial_j$ is accompanied by Clifford algebra representatives $e_j=\mp i\sigma_j$ satisfying $e_je_k+e_ke_j=-2\delta_{jk}\mathbf{1}$. Intrinsically, the two choices are labelled by the \emph{chirality}, $-e_1e_2e_3=\pm\mathbf{1}$, and $H_+$ (resp.\ $H_-$) is the right-handed (resp.\ left-handed) Weyl operator. The possibility of distinct chirality sectors occurs in each odd spatial dimension.

In such local expressions, it is important that the $\{\partial_j\}_{j=1,2,3}$ provide an \emph{oriented orthonormal} tangent frame, in order for $H_\pm$ to enjoy spin-rotation invariances (compare Section \ref{sec:invariant.description}). The symbol of $H_\pm$ is $\pm\sum_{j=1}^3 p_j\sigma_j=\pm\mathbf{p}\cdot\sigma$. In comparison, for a Weyl semimetal, we have \mbox{$\mathbf{h}(\mathbf{k})\cdot\sigma$}, for quasimomentum coordinates $\mathbf{k}=(k_1,k_2,k_3)$. For a non-degenerate Weyl point $w$ with index $\pm 1$ (see Eq.\ \eqref{eqn:non.degenerate.index}), a suitable affine-linear, orientation-preserving, but generally \emph{non-orthogonal} change of coordinates will bring it to the approximate form $\pm \mathbf{k}^\prime\cdot\sigma$ near $w$. 

Thus, provided the metric is adjusted accordingly, a Weyl semimetal behaves near a non-degenerate $w$, and at low energies, like a Weyl fermion. However, it significantly departs from a ``true'' relativistic Weyl fermion once the ``true'' Brillouin torus geometry comes into play. Continuum differential operator models of Weyl semimetals have been investigated in \cite{Witten, Thiang-sf}.

\subsection{Berry connection}
Let us write $X=\widehat{\Gamma}\setminus W$. Recall that the semimetal line bundle $\mathcal{L}_1\to X$ arises as a subbundle of some ambient $X\times \CC^m$, which we assume is \emph{trivialized}. We mention that in the setting of adiabatic perturbation theory, where Berry's notion of geometric phase was initially discussed \cite{Berry}, the state space $\CC^m$ is fixed, independently of the parameter manifold $X$. 

There is a ``trivial'' covariant derivative $\nabla^{\rm triv}$ on $X\times\CC^m$, namely, the ordinary derivative of $\CC^m$-valued \emph{functions} on $X$. The Berry connection on the subbundle $\mathcal{L}_1$ is defined to be
\begin{equation}
\nabla^{\rm Berry}=p\circ \nabla^{\rm triv}\circ \iota,\label{eqn:Berry.connection}
\end{equation}
where $\iota:\mathcal{L}_1\to X\times\CC^m$ is the inclusion and $p:X\times \CC^m\to\mathcal{L}_1$ the orthogonal projection back onto $\mathcal{L}_1$. Simply put, a section $s:X\to\mathcal{L}_1$ is regarded as an ordinary $\CC^m$-valued function, differentiated as such, then projected to a section of $\mathcal{L}_1$.

Here, it is instructive to be precise with terminology. Like any connection, the Berry connection \eqref{eqn:Berry.connection} is gauge-independent and globally defined. What is gauge-dependent is its \emph{local description} as a ${\rm u}(1)$-valued ``connection 1-form'', which of course requires working in some local trivialization (i.e., local gauge choice). Concretely, one usually uses local coordinates $\mathbf{k}=(k_1,k_2,k_3)$ on $\widehat{\Gamma}$ and a (normalized) local section $\mathbf{k}\mapsto|\psi_{\mathbf{k}}\rangle$ of $\mathcal{L}_1$, then writes
\[
\mathcal{A}_j(\mathbf{k})=\langle\psi_{\mathbf{k}}|\tfrac{\partial}{\partial k_j}|\psi_{\mathbf{k}}\rangle,\qquad j=1,2,3,
\]
for the components of the (locally-defined) connection 1-form. There is typically an extra factor of $i$ in physics conventions. 

The \emph{Berry curvature}, $\mathcal{F}=d\mathcal{A}$, is a globally-defined and gauge-independent 2-form over $X$. It can be integrated over closed 2-submanifolds of $X$ to get the various Chern numbers of the Weyl semimetal. 

\begin{example}\label{ex:tautological.connection}
Recall from Section \ref{sec:Bloch.tautological} that $\mathcal{L}^+\cong\mathcal{L}_{\rm taut}$ is a subbundle of $S^2\times \CC^2$. Away from the south pole, we can smoothly choose the normalized vectors
\[
\psi^+(\theta,\varphi)=\binom{\cos\frac{\theta}{2}}{\sin\frac{\theta}{2} e^{i\varphi}}\in \mathcal{L}^+_{(\theta,\varphi)},\qquad\theta\neq\pi
\]
as the local gauge. Then the Berry connection is represented (away from the south pole) as the $\mathfrak{u}(1)$-valued 1-form
\begin{equation}
\mathcal{A}(\theta,\varphi)=\langle\psi^+|d\psi^+\rangle|_{(\theta,\varphi)}=i\sin^2\tfrac{\theta}{2}\,d\varphi,\label{eqn:Bloch.Berry.connection}
\end{equation}
with curvature $\mathcal{F}=\tfrac{i}{2}\sin\theta\,d\theta\wedge d\varphi$. The Chern number of $\mathcal{L}^+$ over $S^2$ is
\[
\frac{i}{2\pi}\int_{S^2}\mathcal{F}=-1.
\]
This is precisely what we found in Section \ref{sec:magnetic.monopole} for the monopole line bundle with $g=-1$. A similar calculation for the bundle $\mathcal{L}^-$ of $(-1)$-eigenspaces relates it to the monopole line bundle with $g=+1$.
\end{example}

As mentioned, the notion of Berry connection and curvature requires the data of an embedding into an ambient trivialized bundle. Whether or not this data is canonically available is a subtle question; see \cite{Moore} for a detailed discussion.

\section{Surface Fermi arcs}
The topological Chern numbers of Weyl semimetals are physically manifested as so-called \emph{Fermi arcs} on the sample surface. The informal argument for this \cite{WTVS} invokes the idea of \emph{bulk-boundary correspondence}. A rigorous derivation based on the idea of topological spectral flow can be found in Section 5 of \cite{Thiang-sf}, and is summarized below.

For ease of discussion, let us assume that the eigenvalue crossings are at $0$-energy, and that $H(\chi)$ has no $0$-eigenvalue when $\chi\not\in W$. For 2-band models, Eq.\ \eqref{eqn:generic.2x2}, this could be achieved by setting $h_0\equiv 0$.

It is assumed that the sample is terminated at a surface parallel to a rank-2 sublattice $\Gamma_\parallel\subset\Gamma$, and there is a splitting
\[
\Gamma\cong \Gamma_\parallel\times\ZZ.
\]
The $\ZZ$ factor is generated by a translation $S_\perp$ transverse to the terminating surface. The Brillouin torus also splits as
\[
\widehat{\Gamma}\cong\widehat{\Gamma_\parallel}\times\TT_\perp,
\]
where we added a subscript $\perp$ to $\TT$ for emphasis. Accordingly, a quasimomentum $\chi\in\widehat{\Gamma}$ is written as
\[
\chi=(\chi_\parallel,\chi_\perp).
\]
We also write $\pi:\widehat{\Gamma}\to\widehat{\Gamma_\parallel}$ for the projection onto the ``surface Brillouin 2-torus'', so $\pi(W)$ is the set of ``projected Weyl points''.

\subsection{Family of Toeplitz operators}
Let $\check{H}$ be the truncation of the tight-binding Hamiltonian $H$ to the half-lattice Hilbert space,
\[
\ell^2(\Gamma_\parallel\times\NN;\CC^m).
\]
Concretely, whenever $S_\perp$ appears in $H$, it is replaced by the unilateral shift $\check{S}_\perp$ in $\check{H}$. Then $\check{H}$ describes a Weyl semimetal occupying a half-space.

Note that $\check{H}$ still commutes with $\Gamma_\parallel$, so we may carry out a partial Fourier transform
\[
\ell^2(\Gamma_\parallel\times\NN;\CC^m)\cong L^2(\widehat{\Gamma_\parallel})\otimes \ell^2(\NN)\otimes\CC^m,
\]
turning $\check{H}$ into a family of self-adjoint operators $\{\check{H}(\chi_\parallel)\}_{\chi_\parallel\in\widehat{\Gamma_\parallel}}$ acting on $\ell^2(\NN)\otimes\CC^m$.

Now, $\ell^2(\NN)\subset\ell^2(\ZZ)$ is, after Fourier transform, the classical Hardy space $H^2(\TT)\subset L^2(\TT)$. So the operator $\check{H}(\chi_\parallel)$ is just the \emph{Toeplitz} operator with smooth symbol function
\[
\TT_\perp\ni \chi_\perp\mapsto H(\chi_\parallel,\chi_\perp)\in{\rm Herm}_0(2).
\]
The essential spectrum (informally the ``bulk spectrum'') of $\check{H}(\chi_\parallel)$ is
\begin{equation}
{\rm ess}\textit{-}{\rm spec}(\check{H}(\chi_\parallel))=\bigcup_{\chi_\perp\in\TT_\perp}{\rm Spec}(H(\chi_\parallel,\chi_\perp)),\label{eqn:ess.spec.Toeplitz}
\end{equation}
see Section 4 of \cite{Arveson}.

\subsection{Loops of self-adjoint Fredholm Toeplitz operators}
If $\chi_\parallel$ is not a projected Weyl point, then $H(\chi_\parallel,\chi_\perp)$ never has $0$ as an eigenvalue, so Eq.\ \eqref{eqn:ess.spec.Toeplitz} says that $0$ does not lie in the essential spectrum of $\check{H}(\chi_\parallel)$. In other words, $\check{H}(\chi_\parallel)$ is an element of $\mathcal{F}^{\rm sa}_*$, the space of self-adjoint \emph{Fredholm} operators possessing both positive and negative essential spectrum. Generally, $\check{H}(\chi_\parallel)$ will also have some discrete spectrum (finite multiplicity eigenvalues) inside this essential spectral gap.

\begin{figure}
\centering
\begin{tikzpicture}
\draw[->] (-0.1,0) -- (4.3,0);
\node [left] at (0.4,1.7) {{\footnotesize{${\rm spec}(\check{H}(\ell(z)))$}}};
\node [left] at (-0.10,0.7) {$+1$};
\node [left] at (-0.10,-0.7) {$-1$};
\node [left] at (-0.10,0) {$0$};
\node [right] at (4.3,0) {$z\in S^1$};
\draw[-] (-0.1,0.7) -- (0.1,0.7);
\draw[-] (-0.1,-0.7) -- (0.1,-0.7);
\filldraw[lightgray] (0,0.7) rectangle (4,1.2);
\filldraw[lightgray] (0,-1.2) rectangle (4,-0.7);
\node [left] at (2,0.9) {${\rm ess}\textit{-}{\rm spec}$};
\node [left] at (2,-1) {${\rm ess}\textit{-}{\rm spec}$};
\draw[thick,domain=2:4] plot (\x,{0.7*cos(0.5*\x*pi r)});
\draw[->] (0,-1.2) -- (0,1.4);
\end{tikzpicture}
\caption{Along a loop $\ell$ in the surface Brillouin torus avoiding the projected Weyl points, the Toeplitz operators $\check{H}(\ell(z))$ are always Fredholm. There may be a net flow of discrete eigenvalues across their common essential spectral gap.}\label{fig:spectral.flow}
\end{figure}
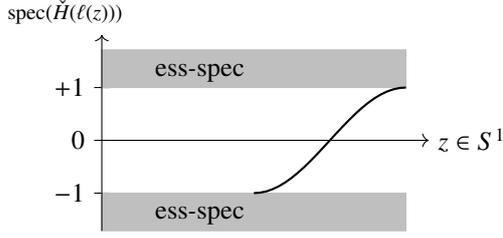

To summarize, we have a norm-continuous map
\[
\check{H}:\widehat{\Gamma_\parallel}\setminus\pi(W)\to \mathcal{F}^{\rm sa}_*.
\]
Any closed loop $\ell:S^1\to \widehat{\Gamma_\parallel}\setminus\pi(W)$ determines a corresponding loop $\check{H}\circ\ell$ of operators in $\mathcal{F}^{\rm sa}_*$. Remarkably, the homotopy class of such an operator loop exactly corresponds to the \emph{spectral flow} of eigenvalues across 0-energy, as the loop is traversed \cite{AS-skew,Phillips}, see Fig.\ \ref{fig:spectral.flow}.

\subsection{Spectral flow and Fermi arcs}
Crucially, a non-trivial spectral flow guarantees that a $0$-energy state must occur somewhere along the loop. Assuming that the Fermi energy is at $0$, the locus in $\widehat{\Gamma_\parallel}$ where $0$-energy states occur is called the \emph{Fermi arc} of the Weyl semimetal. Thus, \emph{the Fermi arc is transverse to those loops in $\widehat{\Gamma_\parallel}\setminus \pi(W)$ which exhibit spectral flow}.

Since spectral flow is a homotopy invariant property of the loop, it will suffice to analyze some standard loops in $\widehat{\Gamma_\parallel}\setminus \pi(W)$. For example, in Figure \ref{fig:two.Weyl.points}, the 2-tori $T_L, T_R$ project onto homotopically distinct loops $\ell_L,\ell_R$. For $T_L$, say, the Bloch Hamiltonians $\{H(\chi)\}_{\chi\in T_L}$ can be viewed as those of a fictitious 2D model Hamiltonian with a spectral gap at $0$-energy. Then, with the help of $K$-theory methods, see Section 5.3 of \cite{Thiang-sf}, one finds that Chern number on $T_L$ corresponds to the spectral flow of $\check{H}$ along $\ell_L$. This is basically the 2D bulk-boundary correspondence. Similarly for $T_R$. 

The conclusion is that the Fermi arc locus in $\widehat{\Gamma_\parallel}$ is the projection of the Dirac string in $\widehat{\Gamma}$, \emph{at the level of relative homology classes}. See also \cite{Gomi} for another derivation of such a result in special models. Thus, the Fermi arc topology is completely determined by the bulk Chern class of the Weyl semimetal. The precise geometric Fermi arc, however, depends on finer details of the Hamiltonian.

\section{Generalizations}
If spatial inversion $\mathsf{P}$ preserves the lattice $\Gamma$, it makes sense to require $H$ to be $\mathsf{P}$-invariant. Momentum is likewise reversed under $\mathsf{P}$, so there is an induced $\ZZ_2$-action on $\widehat{\Gamma}$, usually written in quasimomentum coordinates as $\mathbf{k}\mapsto-\mathbf{k}$. When $\mathsf{P}$-symmetry is present, the Bloch Hamiltonians $H(\mathbf{k})$ and $H(-\mathbf{k})$ are unitarily equivalent, so Weyl points occur in inversion-related pairs, with opposite local indices. The model in Example \ref{ex:minimal.example} is symmetric under $\mathsf{P}=\sigma_3\circ(\mathbf{k}\mapsto-\mathbf{k})$. 

Another possible symmetry is fermionic time-reversal $\mathsf{T}$, which is an antiunitary effecting $\mathbf{k}\mapsto-\mathbf{k}$ and squares to $-1$. This imposes ``fermion doubling'' at $\mathbf{k}=(0,0,0)$ (essentially due to the quaternionic structure imposed there). So models of $\mathsf{T}$-invariant Weyl semimetals require at least four bands, and are more complicated to analyze mathematically \cite{TSG}.

A combined symmetry $\mathsf{P}\mathsf{T}$ would force every Bloch Hamiltonian $H(\chi)$ to respect a quaternionic structure, and have doubly-degenerate eigenvalues. In this case, in a minimal traceless 4-band model, one can write $H(\chi)=\mathbf{h}(\chi)\cdot\gamma$ for suitable gamma matrices $\gamma=(\gamma_1,\ldots\gamma_5)$ (\S4 of \cite{MT}). Similar to the 2-band Weyl semimetal, we get four-fold degenerate eigenvalue crossings whenever $\mathbf{h}(\chi)=0$. This is the setting of a \emph{Dirac semimetal}, which is supposed to be analogous to the relativistic Dirac fermion. This type of eigenvalue crossing is not topologically protected in dimension $d=3$, but it is in dimension $d=5$, with the local index related to the second Chern class of the (rank-2) lower-energy vector bundle, and the Fermi arc phenomenon deducible from higher-dimensional analogues of spectral flow \cite{CT}. To get ``topological protection'' of Dirac semimetals in $d=3$, one needs to invoke other mechanisms, such as extra crystalline symmetries, see \cite{AMV,Gao,Lv}.

Finally we sketch a few other generalizations considered in the physics literature, reviewed in more detail in \cite{AMV,Lv,Gao}. Degenerate Weyl points with quadratic or higher dispersion relations can be considered (``Multi-Weyl'' semimetals). Crystalline symmetry may force band crossings to occur not just at isolated points, but on invariant submanifolds (``nodal lines''). Weyl points need not share exactly the same energy, due to the trace-ful term $h_0\mathbf{1}$. Suppose $h_0\mathbf{1}$ deviates from a constant by an amount larger than the size of the traceless term $H-h_0\mathbf{1}$. Then even if the Weyl points' energies coincide with the Fermi energy $E_F$, the Fermi surface (the level set $\{\chi\in\widehat{\Gamma}\,:\,E_F\in{\rm Spec}(H(\chi))\}$) could comprise a whole surface connecting the Weyl points (``Type II'' Weyl semimetals).

%% The Appendices part is started with the command \appendix;
%% appendix sections are then done as normal sections
%% \appendix

%% \section{}
%% \label{}

%% If you have bibdatabase file and want bibtex to generate the
%% bibitems, please use
%%
%%  \bibliographystyle{elsarticle-num} 
%%  \bibliography{<your bibdatabase>}

%% else use the following coding to input the bibitems directly in the
%% TeX file.

\end{document}